\documentclass[aps,pra,twocolumn,10pt]{revtex4-1}

\usepackage{amsmath,amssymb,bm}
\usepackage{dsfont}
\usepackage{graphics}

\newcommand{\im}{{i}}
\newcommand{\tr}{\text{tr}}

\newcommand{\ket}[1]{\left|{#1}\right\rangle}
\newcommand{\bra}[1]{\left\langle{#1}\right|}

\newcommand{\Dcirc}{{\cal D}^{\circ}}

\newcommand{\eqtri}{\triangleq}

\begin{document}

\title{Nonlinear interferometry in all spatiotemporal degrees of freedom}
\author{Filippus S. \surname{Roux}}
\email{froux@nmisa.org}
\affiliation{National Metrology Institute of South Africa, Meiring Naud{\'e} Road, Brummeria 0040, Pretoria, South Africa}

\begin{abstract}
The effects of the spatiotemporal degrees of freedom on the practical implementation of an SU(1,1) interferometer is investigated. A recently developed Wigner functional approach is used to obtain the phase sensitivity of such an SU(1,1) interferometer in terms of all the spatiotemporal degrees of freedom. It reveals how experimental scale parameters affect the performance of the interferometer. The analysis provides information that would be useful for quantum metrology applications.
\end{abstract}

\maketitle

\section{\label{intro}Introduction}

Quantum metrology can achieve a precision better than the standard quantum limit \cite{maccone}. However, the preparation of the quantum states required in such systems (such as bright squeezed states or N00N states with a large $N$) is often a challenging task. One way to avoid this challenge is with the aid of nonlinear interferometry, also called SU(1,1) interferometry \cite{yurke}.

There are many variants of the SU(1,1) interferometer \cite{su11b,plick,su11seed,su11in1,su11bec}. The initial proposal \cite{yurke} called for a system consisting of two nonlinear crystals where the first produces spontaneous parametric down-converted light, which passes through the second to produce stimulated parametric down-conversion. The latter is then used to measure an observable that is sensitive to a phase modulation applied to the spontaneous parametric down-converted light prior to entering the second crystal. The system sensitivity can be improved by stimulating the light from the first crystal with a coherent state seed \cite{plick,su11seed}. Subsequently, other types of states have been considered for the seed \cite{su11in1} and various other proposals have been made to improve the sensitivity (see for example \cite{su11bec}). The nonlinear SU(1,1) interferometer has been used in different applications \cite{su11in2,oamsquappl}.

Here, we investigate the effects of the spatiotemporal degrees of freedom, as imposed by a bulk optics implementation of an SU(1,1) interferometer. We choose to consider the bulk optics implementation because the spatiotemporal degrees of freedom play a more prominent role is such systems. Although we'll initially derive the expressions of the phase sensitivity for both seeded and unseeded cases, we'll focus on the former in the detailed calculations due to its better performance. For this purpose, we consider a coherent state seed field. The analysis will show how it can be generalized to address other types of quantum states used as seed.

There are several dimension parameters in the practical implementation of such a system that can play a role. Prominent among them are the scale parameters that govern the transverse spatial dimensions and the bandwidth of the frequency spectra. Therefore, we'll focus on the effect of these quantities in our final analysis. Nevertheless, our analysis can also be used to investigate the effects of any other experimental parameters.

\begin{figure}[ht]
\centerline{\includegraphics{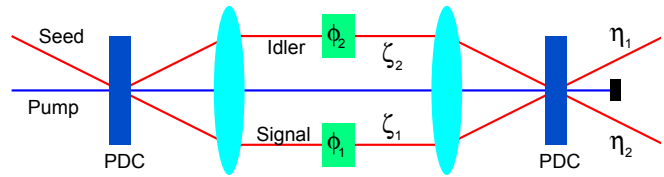}}
\caption{Diagrammatic representation of the seeded SU(1,1) interferometer system.}
\label{syst}
\end{figure}

The output after the first crystal is often regarded as a twin-beam state, even when it is produced by spontaneous parametric down-conversion. The implied ``two beams'' are redirected as input to the second crystal, often with the aid of mirrors. Although stimulated parametric down-conversion produces two distinct beams --- the signal and idler beams --- after the first crystal, spontaneous parametric down-conversion  produces a cone of light with a cone angle determined by the phase-matching conditions. To form two beams out of this cone, one needs to impose a form of post-processing (as implied by the mirrors), which could affect the performance of the system. Although the effect of such post-processing can be readily incorporated in our analysis, it is not our intention to investigate the effect of such post-processing here. Therefore, we consider the situation where the light after the first crystal is guided into the next crystal with the aid of a 4$f$ system (see Fig.~\ref{syst}), ignoring any aperturing effects. The implied $180^{\circ}$ rotation of the transverse plane does not have any effect on the process.

Our analysis is based on a Wigner functional approach \cite{entpdc,queez} incorporating the spatiotemporal degrees of freedom with the particle-number degrees of freedom \cite{stquad,mrowc}. It leads to a functional phase space that generalizes the Moyal formalism \cite{groenewold,moyal,psqm} and represents all quantum optical states without any enforced truncations or approximations. We'll use the results of an investigation of stimulated parametric down-conversion where we incorporated the spatiotemporal degrees of freedom with this Wigner functional approach \cite{paramp}. In the seeded case, our analysis of the effect of the scale parameters produces results reminiscent of those obtained from the investigation of the spatiotemporal effects in the measurement of the squeezing parameter \cite{queez}.

To aid our calculations, we use the {\em thin-crystal approximation}, which is generally a well-satisfied condition in most experimental implementations of parametric down-conversion. The thin-crystal approximation assumes that the Rayleigh range of the pump beam is much larger than the thickness of the nonlinear crystals. The crystal thickness divided by the Rayleigh range provides a dimensionless expansion parameter for the kernel functions. In the {\em thin-crystal limit}, this expansion parameter is set equal to zero, leaving only the leading order term.

The performance of the nonlinear SU(1,1) interferometer is often compared to that of a Mach-Zehnder interferometer. The phase sensitivities of the common phase in the SU(1,1) system is compared to that of the relative phase in the Mach-Zehnder interferometer. The latter saturates the standard quantum limit (or shot-noise limit) under ideal circumstances. The former surpasses the standard quantum limit, approaching the Heisenberg limit, provided that the number of photons that is used in the comparison is taken to be the number of photons ``inside the interferometer,'' making the SU(1,1) system appear more favorable. Such a comparison may be misleading, because the nonlinear process is extremely inefficient and generally uses far fewer photons in the interference process than what is available in the pump. It would be difficult for the SU(1,1) system to beat the performance of the Mach-Zehnder interferometer if the latter is allowed to use the same pump as its input. A comparison between the Mach-Zehnder interferometer and the SU(1,1) interferometer also depends on the nature of the implementation of the SU(1,1) system. In an unseeded system, the number of photons inside the interferometer is spontaneously generated in the first crystal, but in a seeded system the seed is amplified by the nonlinear process to determine the number of photons inside the interferometer. The number of seed photons appears as an additional experimental parameter, leading to a difference in the number of photons that can take part in the interference. While the Mach-Zehnder interferometer is a linear process, the seeded SU(1,1) system is a nonlinear system that amplifies the number of input photons in the seed. Instead of the usual comparison, we ask: how do the experimental parameters affect the improvement in performance due to the amplification in the seeded SU(1,1) system over that of the Mach-Zehnder interferometer, given the same number of input photons? For this reason, the number of seed photons is equated to the number of input photons in the Mach-Zehnder interferometer.

\section{Wigner functional calculation}

\subsection{The output state}

The output of a stimulated parameteric down-conversion process is given by the Bogoliubov transformation of the seed state. For the SU(1,1) interferometer, this Bogoliubov transformation is performed twice. The SU(1,1) interferometer system is diagrammatically represented in Fig. \ref{syst}. For a coherent state seed, the Wigner functional of the output state has the form of a displaced squeezed vacuum state, given by \cite{paramp}
\begin{align}
W_{\text{out}}[\alpha] = & \mathcal{N}_0 \exp\left[-2(\alpha^*-\eta^*)\diamond A\diamond(\alpha-\eta) \right. \nonumber \\
& -(\alpha-\eta)\diamond B^*\diamond(\alpha-\eta) \nonumber \\
& \left.-(\alpha^*-\eta^*)\diamond B\diamond (\alpha^*-\eta^*) \right] ,
\label{kohbogo2}
\end{align}
where $\mathcal{N}_0$ is a normalization constant, $\alpha$ is the field variable that parameterizes the functional phase space on which the Wigner functional is defined, $\eta$ is the parameter function for the displacement of the output state, and $A$ and $B$ are squeezed state kernels produced by both Bogoliubov transformations. The $\diamond$-contraction is a short hand notation for an integration over wave vectors:
\begin{equation}
\alpha^*\diamond A\diamond\alpha \equiv \int \alpha^*(\mathbf{k}) A(\mathbf{k},\mathbf{k}') \alpha(\mathbf{k}')\ \frac{d^2 k d\omega}{(2\pi)^3}\ \frac{d^2 k' d\omega'}{(2\pi)^3} .
\label{binnespek}
\end{equation}
The parameter function of the original input coherent state is given by the twice Bogoliubov transformed parameter function of the output state. In other words, the parameter function for the displaced squeezed vacuum state after the first crystal is given by
\begin{align}
\begin{split}
\zeta = & U_2\diamond \eta + V_2\diamond \eta^* , \\
\zeta^* = & \eta^*\diamond U_2^{\dag} + \eta\diamond V_2^{\dag} ,
\end{split}
\label{zetaeta}
\end{align}
where $U_2$ and $V_2$ are the Bogoliubov kernels of the second nonlinear crystal. The input coherent state parameter function is then given by
\begin{align}
\begin{split}
\xi = & U_1\diamond \zeta + V_1\diamond \zeta^* , \\
\xi^* = & \zeta^*\diamond U_1^{\dag} + \zeta\diamond V_1^{\dag} ,
\end{split}
\label{xizeta}
\end{align}
where $U_1$ and $V_1$ are the Bogoliubov kernels of the first nonlinear crystal. The squeezed state kernels for the combined process are given by
\begin{align}
\begin{split}
A = & U_2\diamond A_1\diamond U_2 + V_2\diamond A_1^T\diamond V_2^* \\
& + U_2\diamond B_1\diamond V_2^* + V_2\diamond B_1^*\diamond U_2 , \\
B = & U_2\diamond A_1\diamond V_2 + V_2\diamond A_1^T\diamond U_2^* \\
& + U_2\diamond B_1\diamond U_2^* + V_2\diamond B_1^*\diamond V_2 ,
\end{split}
\label{abkerne}
\end{align}
where
\begin{align}
\begin{split}
A_1 = & U_1\diamond U_1 + V_1\diamond V_1^* , \\
B_1 = & U_1\diamond V_1 + V_1\diamond U_1^* .
\end{split}
\label{ab1kerne}
\end{align}
Here, it is assumed that in both the Bogoliubov transformations $U$ is Hermitian $U^{\dag}=U$, and $V$ is symmetric $V^T=V$. In addition to these properties, the thin-crystal approximation also implies that
\begin{align}
\begin{split}
U\diamond U - V\diamond V^* = & \mathbf{1} , \\
U\diamond V - V\diamond U^* = & 0 ,
\end{split}
\label{uvident}
\end{align}
where $\mathbf{1}$ represents the identity kernel. With the aid of Eq.~(\ref{ab1kerne}), we can express the combined kernels as
\begin{align}
\begin{split}
A = & A_0^{\dag}\diamond A_0 + B_0^T\diamond B_0^* , \\
B = & A_0^{\dag}\diamond B_0 + B_0^T\diamond A_0^* ,
\end{split}
\label{ab2kerne}
\end{align}
where
\begin{align}
\begin{split}
A_0 = & U_1\diamond U_2 + V_1\diamond V_2^* , \\
B_0 = & U_1\diamond V_2 + V_1\diamond U_2^* ,
\end{split}
\end{align}
which are in general neither symmetric nor Hermitian.

\subsection{Inverse Bogoliubov}

The relationships among the parameter functions are given as input parameter functions in terms of the Bogoliubov transformations of the output parameter functions. From Eqs.~(\ref{zetaeta}) and (\ref{xizeta}), the initial parameter function is related to the final parameter function by
\begin{align}
\begin{split}
\xi = & A_0\diamond \eta + B_0\diamond \eta^* , \\
\xi^* = & \eta^*\diamond A_0^{\dag} + \eta\diamond B_0^{\dag} .
\end{split}
\end{align}
While the $A_0$ operation retains the location of the beam, the $B_0$ operation flips it to the other side. Therefore, we can separate $\eta$ into the two beams as $\eta=\eta_1-\eta_2$ (see Fig.~\ref{syst}). However, $\xi$ is nonzero on only one side. Hence,
\begin{align}
\begin{split}
\xi = & A_0\diamond \eta_1 - B_0\diamond \eta_2^* , \\
0 = & A_0\diamond \eta_2 - B_0\diamond \eta_1^* , \\
\xi^* = & \eta_1^*\diamond A_0^{\dag} - \eta_2\diamond B_0^{\dag} , \\
0 = & \eta_2^*\diamond A_0^{\dag} - \eta_1\diamond B_0^{\dag}  .
\end{split}
\label{xieta}
\end{align}

For the subsequent calculation, we need the output parameter functions $\eta$ in terms of the inverse Bogoliubov transformations of the input parameter functions $\xi$. To invert the transformations, we use the relationships
\begin{align}
\begin{split}
A_0^{\dag}\diamond A_0 - B_0^T\diamond B_0^* = & \mathbf{1} , \\
A_0^{\dag}\diamond B_0 - B_0^T\diamond A_0^* = & 0 ,
\end{split}
\label{ab0vgl}
\end{align}
that follow from Eq.~(\ref{uvident}). These operations then produce the required inverses
\begin{align}
\begin{split}
\eta_1 = & A_0^{\dag}\diamond \xi , \\
\eta_2 = & B_0^T\diamond \xi^* .
\end{split}
\label{etaxi}
\end{align}

\subsection{Phase modulation}

In the SU(1,1) interferometer, the two beams that pass through the $4f$ system are modulated by arbitrary phases to produce an interference effect in the output. The phase modulation is a purely spatial operation that is performed on the parameter function of the state. It can be introduced in the filter plane of the $4f$ system where the angular spectrum of the parameter function is represented in coordinate space, and presents the two beams as spatially separated parts of the spectrum. The phase modulation is therefore easily implemented by applying it as different phase modulations on the separate beams.

For the combined parameter function after the first crystal given by $\zeta=\zeta_1-\zeta_2$, the effect of the phase modulation is to produce
\begin{equation}
\zeta \rightarrow \zeta'=\exp(\im\phi_1)\zeta_1-\exp(\im\phi_2)\zeta_2 ,
\label{sqv}
\end{equation}
where we introduce arbitrary phase modulations for the two beams independently (see Fig.~\ref{syst}). For the combined process, represented by both nonlinear crystals, linking the initial parameter function with the final parameter function, we'll simply insert the appropriate phase into the terms where they would contribute. While the $U$ operations retain the location of the beams, the $V$ operations flip them to the other side. In addition, the field is complex conjugated when it is flipped. Hence,
\begin{align}
\begin{split}
\eta_1 = A_0^{\dag}\diamond \xi
\rightarrow & \exp(\im\phi_1) U_2\diamond U_1 \diamond \xi  \\
& + \exp(-\im\phi_2) V_2\diamond V_1^* \diamond \xi , \\
\eta_2 = \xi^*\diamond B_0
\rightarrow & \xi^*\diamond U_1\diamond V_2 \exp(-\im\phi_1)  \\
& + \xi^*\diamond V_1\diamond U_2^* \exp(\im\phi_2) .
\end{split}
\label{etaxi0}
\end{align}
We can represent the phase modulations in terms of a common phase $\phi_0$ and a relative phase $\phi_{\Delta}$, so that
\begin{align}
\begin{split}
\phi_1 = & \phi_0 + \tfrac{1}{2}\phi_{\Delta} , \\
\phi_2 = & \phi_0 - \tfrac{1}{2}\phi_{\Delta} .
\end{split}
\end{align}
The effect is that the relative phase factor $\exp(\im\phi_{\Delta})$ gives a relative phase modulation for the two beams, while the common phase factor $\exp(\im\phi_0)$ produces an interference effect within the kernel functions. Since the two beams do not overlap, the relative phase acts as a global phase for each beam, respectively, and does not produce any effect. Only the common phase has an effect, which is introduced in the kernels. It is a consequence of the conjugation that is incorporated with the $V$ process.

We also need to keep track of the phase of the pump. The solutions of the Bogoliubov kernels show that the global phase of the pump parameter function appears as a phase factor $\exp(\im\varphi_{\text{p}})$ with $V$, but that $U$ is independent of this phase. In the thin-crystal limit, $U$ is real valued and $V=\im\exp(\im\varphi_{\text{p}})V_0$, where $V_0$ is real valued.

\section{Phase sensitivity measurement}

\subsection{Generating function}

To determine the phase sensitivity, we measure the total intensity of the down-converted light in the final output \cite{plick}. It is proportional to the total number of photons, as obtained from the number operator $\hat{n}$. The result is used to compute the phase sensitivity according to
\begin{equation}
\Delta\phi^2 = \frac{\langle\hat{n}^2\rangle-\langle\hat{n}\rangle^2}{\left(\partial_{\phi}\langle\hat{n}\rangle\right)^2} .
\label{fasesens}
\end{equation}

For the purpose of our calculations, we use the generating function for the Wigner functionals of projection operators \cite{toolbox}, generalized to incorporate the spatiotemporal degrees of freedom. It is given by
\begin{equation}
\mathcal{W}_{\hat{P}} = \left(\frac{2}{1+J}\right)^{\tr\{D\}}
\exp\left(-2 \mathcal{J} \alpha^*\diamond D\diamond\alpha \right) ,
\label{wigproj}
\end{equation}
where $D$ represents a detector kernel, $J$ is the generating parameter, and
\begin{equation}
\mathcal{J} \eqtri \frac{1-J}{1+J} .
\end{equation}
The generating function can be used to compute the photon statistics of a state, observed by a photon-number-resolving detector represented by the kernel $D$. The Wigner functional of the projection operator for $n$ photons is obtained by
\begin{equation}
W_{\ket{n}\bra{n}} [\alpha] = \left. \partial_J^n \mathcal{W}_{\hat{P}} \right|_{J=0} .
\end{equation}
We can compute the different moments directly from the generating function. Applying a derivative to the Wigner function in Eq.~(\ref{wigproj}) before setting $J=1$, we get the Wigner function for the number operator
\begin{equation}
\left. \partial_J \mathcal{W}_{\hat{P}} \right|_{J=1} = \alpha^*\diamond D\diamond\alpha-\tfrac{1}{2}\tr\{D\} \equiv W_{\hat{n}} ,
\end{equation}
which produces the expectation value for the number of photons in a state when it is traced with the Wigner function of that state.

For the second moment we apply two derivatives, where the result after the first derivative is multiplied by $J$. After both derivatives, we set $J=1$.

For the current case under consideration, we'll assume that the detection process measures the total power of the output. For this purpose we can set the detector kernel $D=\mathbf{1}$. It implies that $\tr\{D\}=\Omega$, where $\Omega\sim\infty$ represents that cardinality of the functional phase space.

We multiply the generating function in Eq.~(\ref{wigproj}) with the Wigner functional of the output state given in Eq.~(\ref{kohbogo2}), and evaluate the functional integration over $\beta$.
\begin{widetext}
The result is given by
\begin{align}
\mathcal{W}(J) = & \int W_{\text{fin}}[\beta]\ \mathcal{W}_{\hat{P}}[\beta](J)\ \Dcirc[\beta] \nonumber \\
= & \frac{\mathcal{N}_0 2^{\Omega}}{(1+J)^{\Omega}} \int \exp\left[-2(\beta^*-\eta^*)\diamond A\diamond(\beta-\eta)
-(\beta-\eta)\diamond B^*\diamond(\beta-\eta) \right. \nonumber \\
& \left. -(\beta^*-\eta^*)\diamond B\diamond (\beta^*-\eta^*) -2 \mathcal{J} \beta^*\diamond\beta\right]\ \Dcirc[\beta] \nonumber \\
 = & \frac{\mathcal{N}_0 2^{\Omega}}{(1+J)^{\Omega}} \int \exp\left[-2\alpha^*\diamond A\diamond\alpha
 -\alpha^*\diamond B\diamond\alpha^*-\alpha\diamond B^*\diamond\alpha -2 \mathcal{J}(\alpha^*+\eta^*)\diamond(\alpha+\eta)\right]\ \Dcirc[\alpha] ,
\label{gentotal}
\end{align}
where we shift the integration field variable $\beta\rightarrow\alpha+\eta$. The result after the functional integration is
\begin{align}
\mathcal{W}(J) = & \frac{\exp\left( \mathcal{J}^2\text{-terms} -2 \mathcal{J}|\eta|^2 \right)}
{\sqrt{\det\left\{A+\mathcal{J}\mathbf{1}\right\}
\det\left\{A+\mathcal{J}\mathbf{1}-B\diamond\left(A^*+\mathcal{J}\mathbf{1}\right)^{-1}\diamond B^*\right\}}} ,
\label{gentotal1}
\end{align}
where the $\mathcal{J}^2$-terms, which are only required for the calculation of the second moment, are given by
\begin{align}
\mathcal{J}^2\text{-terms} = & \mathcal{J}^2 \eta^*\diamond A^{-1}\diamond \eta
+ \mathcal{J}^2 \left[\eta^*-\eta\diamond (A^*)^{-1}\diamond B^*\right]\diamond A
 \diamond \left[\eta- B\diamond (A^*)^{-1}\diamond \eta^*\right] \nonumber \\
 \approx & \mathcal{J}^2 \left[ 2\eta^*\diamond A\diamond \eta
-\eta^*\diamond B\diamond \eta^*-\eta\diamond B^*\diamond \eta\right] .
\end{align}
\end{widetext}
Here we set all the $\mathcal{J}$'s inside the kernels to zero. Due to the overall factor of $\mathcal{J}^2$, they would only contribute for moments higher than the second moment. Moreover, we assume the output state is pure, allowing us to set
\begin{equation}
[A-B\diamond (A^*)^{-1}\diamond B^*]^{-1} = A .
\label{suiwer}
\end{equation}

\subsection{First moment and phase derivative}

The first moment is obtained from one derivative of Eq.~(\ref{gentotal1}) with respect to $J$. It produces
\begin{equation}
\langle\hat{n}\rangle = \left. \partial_J \mathcal{W}(J) \right|_{J=1} = |\eta|^2+\tfrac{1}{2} \tr\{A-\mathbf{1}\} .
\label{momeen}
\end{equation}
For a strong enough seed field, the first term would completely dominate over the second term, which is the spontaneously parametric down-converted background light. Hence, we can assume
\begin{equation}
\langle\hat{n}\rangle = |\eta|^2 = |\eta_1|^2  + |\eta_2|^2 ,
\label{momeen1}
\end{equation}
where $\eta=\eta_1-\eta_2$ is given in terms of Eq.~(\ref{etaxi0}). On the other hand, in those cases where there is no seed field, the first moment is given by only the spontaneous parametric down-converted light
\begin{equation}
\langle\hat{n}\rangle = \tfrac{1}{2} \tr\{A-\mathbf{1}\} .
\label{momeen2}
\end{equation}

Now, we apply derivatives on the phase $\phi_0$. When Eq.~(\ref{etaxi0}) is substituted into Eq.~(\ref{momeen1}), the relative phase cancels everywhere. Therefore, we apply the derivative with respect to the common phase. It leads to
\begin{align}
\partial_{\phi_0}\langle\hat{n}\rangle = &
 - \im 2\exp(-\im 2\phi_0)\xi^*\diamond U_1\diamond B_2\diamond V_1^*\diamond\xi \nonumber \\
& + \im 2\exp(\im 2\phi_0)\xi^*\diamond V_1\diamond B_2^*\diamond U_1\diamond\xi ,
\label{dpmomeen0}
\end{align}
where
\begin{align}
\begin{split}
B_2 = & U_2\diamond V_2 + V_2\diamond U_2^* , \\
B_2^* = & U_2^*\diamond V_2^* + V_2^*\diamond U_2 .
\end{split}
\end{align}
The total quantity is real valued because the two terms are complex conjugates of each other. Representing the contractions as
\begin{align}
\begin{split}
\xi^*\diamond U_a\diamond B_b\diamond V_a^*\diamond\xi = & G_1 = |G_1| \exp(\im\gamma_1) , \\
\xi^*\diamond V_a\diamond B_b^*\diamond U_a\diamond\xi = & G_1^* = |G_1| \exp(-\im\gamma_1) ,
\end{split}
\label{xibxi}
\end{align}
we obtain the simpler expression
\begin{equation}
\partial_{\phi_0}\langle\hat{n}\rangle = 4|G_1| \sin(2\phi_0-\gamma_1) .
\label{dpmomeen1}
\end{equation}

\subsection{Second moment and the variance}

The second moment is compute by
\begin{align}
\langle\hat{n}^2\rangle = & \left. \partial_J \left[ J \partial_J \mathcal{W}(J) \right] \right|_{J=1}\nonumber \\
= & \eta^*\diamond A\diamond\eta-\tfrac{1}{2}\eta^*\diamond B\diamond\eta^*-\tfrac{1}{2}\eta\diamond B^*\diamond\eta\nonumber \\
& + \|\eta\|^4 + \|\eta\|^2\tr\{A-\mathbf{1}\} \nonumber \\
& + \tfrac{1}{4} \left(\tr\{A-\mathbf{1}\}\right)^2 + \tfrac{1}{2}\tr\{A\diamond A-\mathbf{1}\} ,
\end{align}
where we used Eq.~(\ref{suiwer}) to assume that
\begin{equation}
A\diamond B\diamond (A^*)^{-1}\diamond (A^*)^{-1}\diamond B^* \approx A-A^{-1} .
\end{equation}

The variance is then given by
\begin{align}
\sigma^2 = & \langle\hat{n}^2\rangle-\langle\hat{n}\rangle^2 \nonumber \\
= & \eta^*\diamond A\diamond\eta-\tfrac{1}{2}\eta^*\diamond B\diamond\eta^*-\tfrac{1}{2}\eta\diamond B^*\diamond\eta \nonumber \\
& + \tfrac{1}{2}\tr\{A\diamond A-\mathbf{1}\} .
\end{align}
If we assume that $\eta$ is a strong field, then we can discard the background term and only have
\begin{equation}
\sigma^2 = \eta^*\diamond A\diamond\eta-\tfrac{1}{2}\eta^*\diamond B\diamond\eta^*-\tfrac{1}{2}\eta\diamond B^*\diamond\eta .
\label{vareen}
\end{equation}
For the case without an input seed field $\eta=0$, we have
\begin{equation}
\sigma^2 = \tfrac{1}{2}\tr\{A\diamond A-\mathbf{1}\} .
\label{varagter}
\end{equation}

In terms of Eq.~(\ref{ab2kerne}), the variance in Eq.~(\ref{vareen}) can be represented as
\begin{align}
\sigma^2 = & (\eta_1^*\diamond A_0^{\dag}-\eta_2\diamond B_0^{\dag})\diamond (A_0\diamond\eta_1-B_0\diamond\eta_2^*) \nonumber \\
& + (\eta_1^*\diamond B_0^T-\eta_2\diamond A_0^T) \diamond (B_0^*\diamond\eta_1-A_0^*\diamond\eta_2^*) .
\label{varpsen}
\end{align}
Based on the identities in Eq.~(\ref{uvident}), it follows that
\begin{align}
\begin{split}
A_0\diamond\eta_1-B_0\diamond\eta_2^*
= & \left[\cos(\phi_0) \mathbf{1} - \im\sin(\phi_0) A_1 \right] \diamond \xi , \\
B_0^*\diamond\eta_1-A_0^*\diamond\eta_2^*
= & \im\sin(\phi_0) B_1^*\diamond \xi .
\end{split}
\end{align}
We can now substitute these expressions into Eq.~(\ref{varpsen}), and get
\begin{equation}
\sigma^2 = \cos^2(\phi_0) \|\xi\|^2 + \sin^2(\phi_0) G_0 ,
\label{defvar}
\end{equation}
where we define
\begin{equation}
G_0 \eqtri \xi^*\diamond\left(A_1\diamond A_1 + B_1\diamond B_1^*\right)\diamond\xi .
\label{defg0}
\end{equation}

\subsection{Phase sensitivity expression}

The phase sensitivity in the SU(1,1) interferometer is obtained by substituting Eqs.~(\ref{dpmomeen1}) and (\ref{defvar}) into Eq.~(\ref{fasesens}). The result reads
\begin{equation}
\Delta\phi_0^2 = \frac{\cos^2(\phi_0) \|\xi\|^2 + \sin^2(\phi_0) G_0}{16 |G_1|^2 \sin^2(2\phi_0-\gamma_1)} .
\end{equation}
The quantities $G_0$ and $|G_1|$ depend on the detail calculations involving the kernels and the parameter function of the seed. In general, $\gamma_1$ can be identified with the global phase of the pump at the second crystal relative to the first. Based on the definition of $G_0$ in Eq.~(\ref{defg0}) and the properties of the kernels $A_1$ and $B_1$, it is clear that $G_0\geq \|\xi\|^2$. Hence, to minimize $\Delta\phi_0^2$, we need to consider the point where $\phi_0=0$ and $\gamma_1=\pm\pi/2$. It gives
\begin{equation}
\Delta\phi_{0,\text{min}} = \frac{\|\xi\|}{4 |G_1|} = \frac{\sqrt{N_{\text{s}}}}{4 |G_1|} ,
\end{equation}
where $N_{\text{s}}=\|\xi\|^2$ represents the average number of photons in the seed. In the weak squeezing limit, we have $U_a\sim \mathbf{1}$, $V_a\propto\Xi_a$ and $B_b\propto\Xi_b$. Based on the definition in Eq.~(\ref{xibxi}), we get $|G_1|\sim \Xi_1\Xi_2 N_{\text{s}}$, where $\Xi_{1,2}$ represents the queezing parameters of the two nonlinear processes, respectively, each being proportional to $\sqrt{N_{\text{p}}}$, where $N_{\text{p}}$ is the average number of photons in the pump. Hence,
\begin{equation}
\Delta\phi_0 \sim \frac{1}{4\Xi_a\Xi_b\sqrt{N_{\text{s}}}} \propto \frac{1}{N_{\text{p}}\sqrt{N_{\text{s}}}} .
\end{equation}
It shows that an increase in squeezing gives a reduction in $\Delta\phi_0$. However, for larger squeezing, the behavior becomes that of a hyperbolic sinusoidal function.

For comparison, we consider the phase sensitivity that is obtained in the Mach-Zehnder interferometer with respect to the relative phase $\phi_{\Delta}$. It is given by
\begin{equation}
\Delta\phi_{\Delta}^2 = 2\frac{1+\cos(\phi_{\Delta})}{N_{\text{in}} \sin^2(\phi_{\Delta})} ,
\end{equation}
and is minimized in the limit where $\phi_{\Delta}\rightarrow\pi$, leading to
\begin{equation}
\Delta\phi_{\Delta,\text{min}} = \frac{1}{\sqrt{N_{\text{in}}}} .
\end{equation}
This result coincides with the standard quantum limit.

We see that the SU(1,1) interferometer does not change the behavior with respect to the number of input photons $N_{\text{in}}$, if we associate it with $N_{\text{s}}$. However, it produces a smaller value thanks to the amplifications imposed by the two nonlinear processes. Here, we study the improvement of the seeded SU(1,1) interferometer over the Mach-Zehnder interferometer by considering the ratio
\begin{equation}
\rho \eqtri \frac{\Delta\phi_{0,\text{min}}}{\Delta\phi_{\Delta,\text{min}}} .
\end{equation}
For $N_{\text{in}}=N_s$, the effect is to remove the dependence on $N_s$. The standard quantum limit is then given when this ratio is $\rho=1$.

\section{Detail calculations}

\subsection{Thin-crystal limit}

To get a more precise indication of the improvement, we use the expressions of the kernels in the thin-crystal limit and calculate of the contractions and the overlap with a seed field. For definitiveness, we choose a parametric down-conversion process based on type I phase-matching. However, similar calculations can be done for other scenarios, even when the nonlinear process is based on four-wave mixing. The resulting calculations are similar to those done in \cite{queez}.

The parameter function of the seed field is assumed to be given by a Gaussian angular spectrum
\begin{equation}
\xi = \sqrt{2\pi} \xi_0 w_{\text{s}} \exp\left(-\tfrac{1}{4} w_{\text{s}}^2 |\mathbf{K}|^2 \right)
h(\omega-\omega_{\text{s}};\delta_{\text{s}}) ,
\label{saaddef}
\end{equation}
where $\xi_0=\|\xi\|$ is the magnitude of the function, and $w_{\text{s}}$ is its beam waist radius at the crystal plane. The spectrum is represented by $h(\omega-\omega_{\text{s}};\delta_{\text{s}})$ as a (monochromatic) narrow spectral function with a bandwidth $\delta_{\text{s}}$ and a center frequency $\omega_{\text{s}}$. For the sake of the calculations, we model $h(\omega-\omega_{\text{s}};\delta_{\text{s}})$ as a normalized Gaussian function.

Here, we use the Bogoliubov kernels in the thin-crystal limit, given by
\begin{align}
\begin{split}
U = & \mathbf{1}+\sum_{n=1}^{\infty} \frac{1}{4^n}H_{2n}^{(\text{e})} , \\
V = & \im\exp(\im\varphi)\sum_{n=1}^{\infty} \frac{2}{4^n} H_{2n-1}^{(\text{o})} ,
\end{split}
\label{defuvdkl}
\end{align}
in terms of
\begin{align}
\begin{split}
H_m^{(\text{o})}
= & \im\frac{M_0 M_1^m}{m^{5/4}m!}\omega_1^{m/2}(\omega_{\text{p}}-\omega_1)^{m/2} \\
& \times h(\omega_{\text{p}}-\omega_1-\omega_2,\sqrt{m}\delta_{\text{p}}) \\
& \times \exp\left(-\frac{w_{\text{p}}^2 |\mathbf{K}_1+\mathbf{K}_2|^2}{4m}\right) ~~~ \text{for~odd} ~ m , \\
H_m^{(\text{e})}
= & \frac{M_0 M_1^m}{m^{5/4}m!}\omega_1^{m/2}(\omega_{\text{p}}-\omega_1)^{m/2} \\
& \times h(\omega_1-\omega_2,\sqrt{m}\delta_{\text{p}}) \\
& \times \exp\left(-\frac{w_{\text{p}}^2 |\mathbf{K}_1-\mathbf{K}_2|^2}{4m}\right) ~~~ \text{for~even} ~ m ,
\end{split}
\label{kernedkl}
\end{align}
where $\mathbf{K}$ is the two-dimensional transverse part of the wavevector $\mathbf{k}$, and
\begin{equation}
M_0 = \frac{\pi^{5/4} w_{\text{p}}^2}{\sqrt{\delta_{\text{p}}}} , ~~~
M_1 = \frac{4\sqrt{2}L|\psi_0|\sigma_{\text{ooe}}\sqrt{\omega_{\text{p}}\delta_{\text{p}}}}{\pi^{3/4} c^2 w_{\text{p}}} .
\label{defmm}
\end{equation}
Here $|\psi_0|$ is the magnitude of the pump parameter function, $w_{\text{p}}$ is its beam waist radius at the crystal plane, $\omega_{\text{p}}$ is the center frequency, $\delta_{\text{p}}$ is the bandwidth, $L$ is the length of the nonlinear crystal, $\sigma_{\text{ooe}}$ is the nonlinear coefficient of the crystal for type I phase matching expressed as a cross-section area, and $c$ is the speed of light.

The Bogoliubov kernels are given as summations of contractions $H_m^{o,e}\equiv H_0^{m\diamond}$ of a bilinear kernel $H_0$. It is obtained under the semiclassical approximation with a coherent state pump, parameterized by a Gaussian parameter function, given by
\begin{equation}
\psi = \sqrt{2\pi} \psi_0 w_{\text{p}} \exp\left(-\tfrac{1}{4} w_{\text{p}}^2 |\mathbf{K}|^2 \right)
h(\omega-\omega_{\text{p}};\delta_{\text{p}}) .
\label{pompdef}
\end{equation}

Under the assumption that the conditions at the two crystals are the same, apart from a different global phase for the pump, the calculations lead to
\begin{align}
\begin{split}
G_0 = & \sum_{n=0}^{\infty} \frac{|\xi_0|^2 (2\Xi)^{2n}}{(2n)!(1+n\nu)\sqrt{1+n\mu}} , \\
|G_1| = & \sum_{n=1}^{\infty} \frac{|\xi_0|^2 (2\Xi)^{2n}}{(2n)!(1+n\nu)\sqrt{1+n\mu}} ,
\end{split}
\label{defg0g1}
\end{align}
where
\begin{equation}
\nu = \frac{w_{\text{s}}^2}{w_{\text{p}}^2} , ~~~
\mu = \frac{\delta_{\text{p}}^2}{\delta_{\text{s}}^2} ,
\label{defnuzeta}
\end{equation}
and
\begin{equation}
\Xi = \frac{L|\psi_0|\sigma_{\text{ooe}}\omega_{\text{p}}^{3/2}\sqrt{\delta_{\text{p}}}}{\sqrt{2}\pi^{3/4} c^2 w_{\text{p}}} ,
\label{sqpardef}
\end{equation}
is the squeezing parameter. We see that the only difference in the two quantities in Eq.~(\ref{defg0g1}) is the starting point of the summation. Hence, $|G_1|=G_0-|\xi_0|^2$. The summation for $G_0$ is similar to the one in \cite{queez}. Since we have two nonlinear processes here, the squeezing parameter is multiplied by 2 in the current case. The summation is not tractible in this form. Different approximations were considered in \cite{queez} to study the behavior.

\begin{figure}[ht]
\centerline{\includegraphics{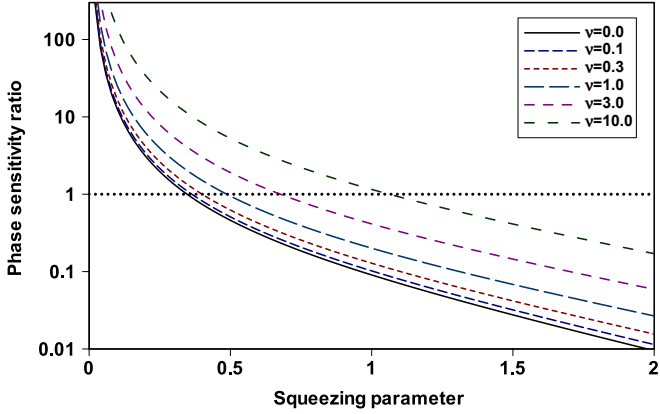}}
\caption{The minimum phase sensitivity ratio $\rho$ is plotted as a function of the squeezing parameter $\Xi$ for different values of $\nu$ with $\mu=0$. The standard quantum limit threshold is indicated by a dotted line.}
\label{nu}
\end{figure}

\subsection{Ideal case}

If we can neglect the factors containing $\nu$ and $\mu$ in Eq.~(\ref{defg0g1}), which would mean that $w_{\text{s}}\ll w_{\text{p}}$ and $\delta_{\text{p}} \ll \delta_{\text{s}}$, then the summations evaluate to
\begin{align}
\begin{split}
G_0 = & |\xi_0|^2 \cosh(2\Xi) , \\
|G_1| = & 2|\xi_0|^2 \sinh^2(\Xi) .
\end{split}
\end{align}
Hence,
\begin{equation}
\Delta\phi_0^2 = \frac{1 + 2 \sin^2(\phi_0) \sinh^2(\Xi)}{64 |\xi_0|^2 \sinh^4(\Xi) \sin^2(2\phi_0-\gamma_1)} .
\end{equation}
Then, for $\phi_0=0$ and $\gamma_1=\pm\pi/2$, the minimum is obtained as
\begin{equation}
\Delta\phi_{0,\text{min}}
 = \frac{1}{8 |\xi_0| \sinh^2(\Xi)}
 = \frac{1}{8 \sqrt{N_{\text{s}}} \sinh^2(\Xi)} .
\end{equation}

\subsection{Finite seed beam width}

Consider now the case where $\mu=0$ and $\nu>0$, as we would have with a finite seed beam width, comparable to the pump beam width. The summation for $|G_1|$ then becomes
\begin{align}
|G_1| = & \sum_{n=1}^{\infty} \frac{|\xi_0|^2 (2\Xi)^{2n}}{(2n)!(1+n\nu)} \nonumber \\
 = & |\xi_0|^2 \left[ {_1F_2}\left(\frac{1}{\nu};\frac{1}{2},1+\frac{1}{\nu};\Xi^2\right)-1\right] ,
\end{align}
where ${_1F_2}$ is a hypergeometric function.

In Fig.~\ref{nu}, we plot the minimum phase sensitivity ratio $\rho$ as a function of the squeezing parameter $\Xi$ with $\mu=0$ for different values of $\nu$, including the ideal case where $\nu=0$. The effect of $\nu$ is to reduce the squeezing, which in turn increases the value of $\rho$. In general, larger values of $\nu$ require larger values of the squeezing parameter $\Xi$ to cross the standard quantum limit threshold, which is indicated by the dotted line in Fig.~\ref{nu}. To achieve the same value for the minimum phase sensitivity ratio at a larger value of $\nu$, we need to increase the amount of squeezing. As a comparison, the amount of squeezing needed for $\nu=1$ when $w_{\text{s}}=w_{\text{p}}$ to achieve the same performance as in the ideal case is about $40\%$ larger. Therefore, the best performance for a given amount of squeezing is obtained for $w_{\text{s}}\ll w_{\text{p}}$.

\begin{figure}[ht]
\centerline{\includegraphics{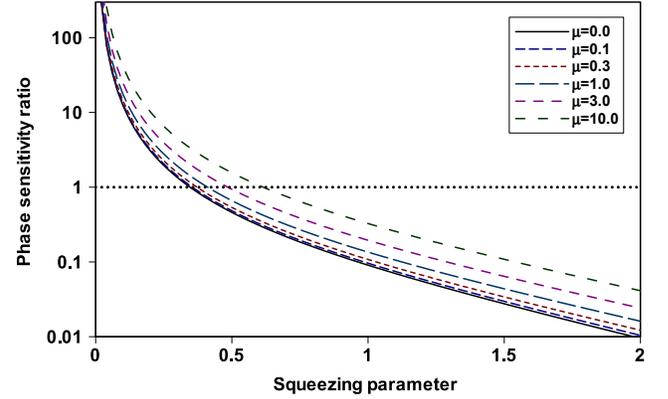}}
\caption{The minimum phase sensitivity ratio $\rho$ is plotted as a function of the squeezing parameter $\Xi$ for different values of $\mu$ with $\nu=0$. The standard quantum limit threshold is indicated by a dotted line.}
\label{mu}
\end{figure}

\subsection{Finite bandwidth}

When we consider the case where $\nu=0$ and $\mu>0$, the resulting summation does not evaluate to a closed form. Here, we use the auxiliary integral \cite{queez},
\begin{equation}
\frac{1}{\sqrt{a}} = \frac{1}{\sqrt{\pi}} \int \exp(-a x^2)\ dx ,
\end{equation}
to replace the factor $\sqrt{1+n\mu}$. The resulting function can be summed so that
\begin{align}
|G_1| = & \sum_{n=1}^{\infty} \frac{|\xi_0|^2 (2\Xi)^{2n}}{(2n)!\sqrt{1+n\mu}} \nonumber \\
 = & \frac{|\xi_0|^2}{\sqrt{\pi}} \int \exp(-x^2) \cosh\left[2\Xi\exp\left(-\tfrac{1}{2}\mu x^2\right)\right]\ dx\nonumber \\
 & - |\xi_0|^2 .
\end{align}
It represents an ensemble average of different values of $|G_1|$ for a varying squeezing parameter given by $\Xi\exp(-\tfrac{1}{2}\mu x^2)$ under a Gaussian probability density. Its effect is to reduce the squeezing, depending on $\mu$. The integral over $x$ cannot be evaluated. However, it can be approximated by a summation over discrete values of $x$. The resulting curves of the minimum phase sensitivity ratio $\rho$ are shown in Fig.~\ref{mu} as a function of $\Xi$ for different values of $\mu$, including the ideal case where $\mu=0$, and with $\nu=0$. The effect of a finite value for $\mu$ is to reduce the squeezing, which increases the value of $\rho$, similar to the case for a finite $\nu$. However, the effect of $\mu$ is less severe as the effect of $\nu$. While, a larger value of $\mu$ requires a larger value of $\Xi$ to cross the standard quantum limit threshold, indicated by the dotted line in Fig.~\ref{mu}, the required increase in $\Xi$ is not as large as with larger values of $\nu$. Here, the amount of squeezing needed for $\mu=1$ when $\delta_{\text{s}}=\delta_{\text{p}}$ to achieve the same performance as in the ideal case is only about $18\%$ larger. Still, the best performance is obtained for $\delta_{\text{s}}\gg\delta_{\text{p}}$.

\subsection{Discussion}

The results of the investigation indicates that the performance of the system is improved for smaller values of $\nu$ and $\mu$. The reason for these trends can be found in the relationship between these quantities and the spatiotemporal information capacity of the system.

Starting with $\nu$, which is the squared ratio of the seed beam width to the pump beam width, we note that the width of the angular spectrum of a beam is inversely proportional to its beam width: a very large beam width represents a narrow angular spectrum. The pump's spectral width serves as the resolution in terms of which the angular spectrum of the seed is resolved. It follows from the fact that the pump governs the parameters in the kernels that mediate the amplification of the seed field. Therefore, if the pump beam's angular spectrum is much smaller than the seed beam's angular spectrum, then a larger number of spatial degrees of freedom can be transferred through the system. The ratio of the seed angular spectral width to the pump angular spectral width can therefore represents a kind of {\em space-bandwidth product}, which in turn represents the capacity of the system to convey spatial information. However, $\nu$ is defined as the inverse of this ratio. Therefore, a smaller value of $\nu$ implies a larger value of the space-bandwidth product, giving a larger spatial information capacity.

A similar understanding follows for $\mu$. In this case, it is related to a kind of {\em time-bandwidth product}, which represents the capacity of the system to convey temporal information, but with a few differences. Firstly, $\mu$ is defined directly in terms of the temporal bandwidths, but with the pump bandwidth on top. Secondly, while the space-bandwidth product represents a two-dimensional space, the time-bandwidth product is associated with a one-dimensional space. So, if the seed beam's temporal spectrum is resolved in terms of the pump spectral width, then $\mu$ is proportional to the inverse square of the time-bandwidth product. A smaller value of $\mu$ thus represents a larger time-bandwidth product, leading to a larger temporal information capacity.

\section{Conclusions}

The effect of the dominant spatiotemporal scale parameters in a practical implementation of a nonlinear SU(1,1) interferometer is investigated. Using the Wigner functional approach, which incorporates all the spatiotemporal degrees of freedom without the need for discretization or truncation, we perform the calculations directly in terms of the kernel functions and thus do not require knowledge of the eigenbasis of the process \cite{sharapova}. It makes it possible to perform calculations to all orders in the expansion of the kernels under the thin-crystal approximation. At the same time, it incorporates all the relevant experimental parameters to make the analytically calculated result relevant for practical experimental investigations. Thus, it allows one to investigate the effects of such experimental parameters with analytical results without the need for numerical simulations.

Under ideal circumstances, which do not correspond to the situation in practical implementations, the phase sensitivity in a nonlinear interferometer follows the optimal trend as determined by the theoretical amount of squeezing. The current analysis shows how this ideal performance is affected by the dominant scale parameters in the implementation. To approach the ideal performance the bandwidth of the seed field needs to much larger than that of the pump field. At the same time, the beam width of the seed field needs to much smaller than that of the pump field. The effect of the beam width is more severe than that of the bandwidth.

\section*{Acknowledgement}

This work was supported in part by funding from the National Research Foundation of South Africa (Grant Number: 118532) and from the Department of Science and Innovation (DSI) through the South African Quantum Technology Initiative.

\end{document}